# Edge Offloading in Smart Grid

**Gabriel Ioan Arcas[1,2], Tudor Cioara[2*], Ionut Anghel[2*], Dragos Lazea[2], Anca Hangan[2]**

[1] Bosch Engineering Center, Bosch Engineering Center, Cluj-Napoca, Romania; gabriel.arcas@ro.bosch.com
[2] Computer Science Department, Technical University of Cluj-Napoca, Memorandumului 28, 400114 Cluj-Napoca, Romania; tudor.cioara@cs.utcluj.ro, ionut.anghel@cs.utcluj.ro; anca.hangan@cs.utcluj.ro, dragos.lazea@cs.utcluj.ro.
* Corresponding author

**Abstract:** The energy transition supports the shift towards more sustainable energy alternatives, paving towards decentralized smart grids, where the energy is generated closer to the point of use. The decentralized smart grids foresee novel data-driven low latency applications for improving resilience and responsiveness, such as peer-to-peer energy trading, microgrid control, fault detection, or demand response. However, the traditional cloud-based smart grid architectures are unable to meet the requirements of the new emerging applications such as low latency and high-reliability thus alternative architectures such as edge, fog, or hybrid need to be adopted. Moreover, edge offloading can play a pivotal role for the next-generation smart grid AI applications because it enables the efficient utilization of computing resources and addresses the challenges of increasing data generated by IoT devices, optimizing the response time, energy consumption, and network performance. However, a comprehensive overview of the current state of research is needed to support sound decisions regarding energy-related applications offloading from cloud to fog or edge, focusing on smart grid open challenges and potential impacts. In this paper, we delve into smart grid and computational distribution architectures, including edge-fog-cloud models, orchestration architecture, and serverless computing, and analyze the decision-making variables and optimization algorithms to assess the efficiency of edge offloading. Finally, the work contributes to a comprehensive understanding of the edge offloading in smart grid, providing a SWOT analysis to support decision making.

**Keywords:** smart grid; edge offloading; edge-cloud integration; offloading criteria, edge orchestration; metaheuristics; reinforcement learning.

## 1. Introduction

As IoT sensors and actuators are deployed in smart grids, the operation and control need real-time processing closer to the edge for faster response and to support the development of context-aware, AI-driven energy services [1]. This trend is accelerated by the renewable energy sources integration at the edge of the grid which requires holistic solutions and decentralized energy and computational infrastructures to assure energy resilience and decrease of carbon footprint [2]. However, in smart grid decentralized scenarios, the offloading of processing workloads towards the edge nodes is challenging due to the heterogeneity, diversity of resources and applications characteristics as well as edge uncertainty [1]. Challenges like real-time data processing, reducing latency, and security need to be systematically addressed in smart grid and edge and fog computing can play a fundamental role for energy sector decentralization [3].

Different edge computing and energy grid-related factors need to be considered to offload and orchestrate in near real-time applications at the edge of the smart grid to address operational problems brought by the integration of renewable energy sources while minimizing the data transfers [1], [4]. The challenge is to make optimal computational orchestration decisions under uncertain and dynamic conditions [5] given by edge resource capacity demand (e.g., bandwidth and memory), failures (e.g., data network link), the latency of the network, energy consumption of resources and lifecycle activities of applications. Automation is a key aspect in managing edge offloading solutions in smart grids and is facilitated by recent advancements in applications virtualization, semantic integration, and data connectivity of edge devices [6].

Moreover, the edge offloading decisions are also influenced by the contextual aspects of the data, encompassing requirements like low response time and various network performance characteristics [7]. Edge AI is emerging as a new paradigm for the efficient management of smart grids, leveraging the improvement of machine learning models that can run at the edge of the grid [8]. It is facilitated by factors such as the development of training pipelines with improved usability, advancements in computing infrastructure at the edge that happen at a higher rate than the reduction of wide area networks latency, and adoption of IoT devices in the smart grid that generates big data that need to be processed and considered by AI [5], [9]. Edge-fog-cloud federated frameworks offer promising solutions for processing data using AI at the edge nodes and orchestrating a global model in the cloud [10], [11]. Nevertheless, their applications in smart grid scenarios and integration with new real-time context-aware energy assets management services are rather limited, even though they bring clear benefits in terms of data management in smart grid, privacy, and security, or addressing latency impact on services' delivery. However, nowadays energy services focus on assuring the links and connectors for analyzing data in the cloud, taking advantage of the potential unlimited computational resources [12].

In this context, a comprehensive overview of the current state of research is needed to support sound decisions regarding smart grid applications offloading from cloud to fog or edge. The edge offloading implementation is complex, requiring substantial upfront investments and posing integration and security challenges. This report aims to bridge knowledge gaps, serving as a comprehensive guide that explores edge offloading in the energy sector, focusing on architecture, criteria, and decision-making techniques. Existing architecture and offloading decision-making criteria need to be analyzed in the context of the smart grid to support applications orchestration across the computing continuum, supporting the implementation and delivery of AI-driven energy services at the edge of the smart grid. We overview the smart grid and computational distribution architectures, including edge-fog-cloud models, orchestration architecture, and serverless computing, considering decentralization and the case of edge offloading. Despite their potential, these architectures face challenges in coordinating tasks due to the complexity of management across layers. As the optimization problem is computationally complex and involves a high dimensionality of the solution space, it is addressed using heuristics-based computing or reinforcement learning models. We analyze the decision-making variables and optimization algorithms to assess their efficiency and applicability to edge offloading. Finally, we provide a SWOT analysis to support edge-offloading decision-making in smart grids, improve computational resource allocation, and enhance overall smart grid decentralized organization.

The rest of the paper is structured as follows. Section 2 presents the basic concepts of edge, fog, and smart grids. Section 3 offers an overview of existing architectures for smart grid and edge offloading. Section 4 analyses the criteria used in offloading decision-making, and Section 5 focuses on heuristics and reinforcement learning solutions. Section 6 concludes the paper and discusses the strengths, weaknesses, opportunities, and threats related to edge offloading in smart grid.

**2. Basic concepts**

The emergence of new IoT technologies and intelligent infrastructure models led to a significant increase in the number of network-connected devices and the volume of data that moves across the network. Consequently, traditional data processing performed entirely in a cloud environment resulted in large communication latencies, making it difficult to deliver real-time results in internet-based applications [13]. These applications run mainly on the end users' mobile devices, which are limited in terms of computational resources and storage capacity, while data processing ensures the functionalities are executed in the cloud. In this context, using the traditional network architecture creates a high network load and communication becomes completely inefficient [14]. Edge and fog computing paradigms emerged to address the bottlenecks of cloud-based architectures by moving the data processing at the edge of the network, closer to the place where it is generated and consumed. Edge and fog computing are important in offloading cloud-based applications by providing the required computational and storage resources and services closer to the users (see Figure 1).

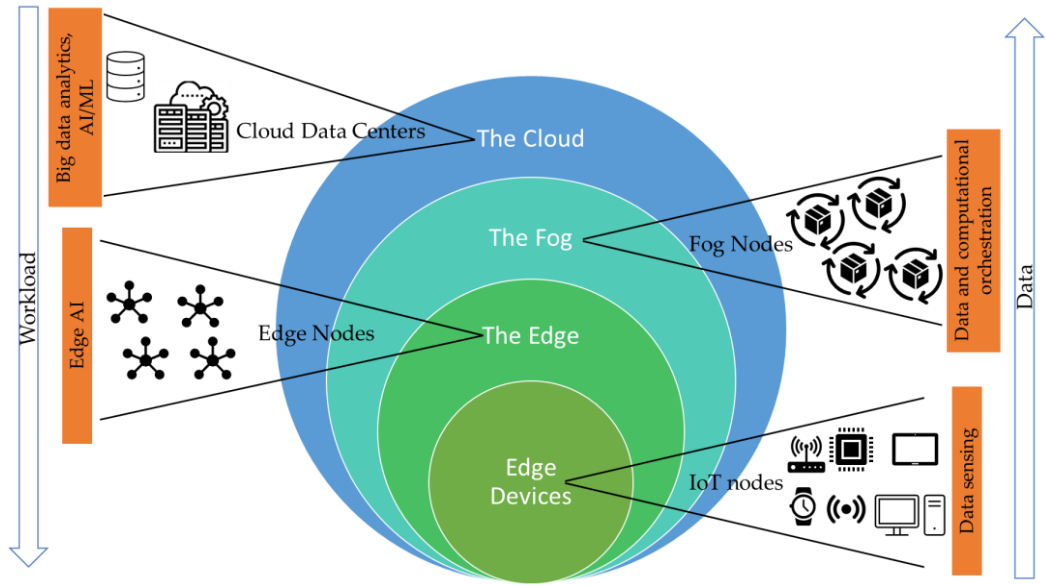

**Figure 1.** The edge-fog-cloud architecture

In the smart grid context, the integration of IoT monitoring devices led to the generation of big data that challenges the nowadays cloud-based applications due to latency and responsiveness problems. In this sense, the edge or fog computing infrastructure could be used between the energy grid monitoring devices and the cloud level, enabling data processing closer to the edge, and reducing the data exchanges with the cloud [15]. Edge servers can be deployed with enough processing capacity to allow the analysis of IoT data and provide faster decision-making for optimizing decentralized energy systems and the data being processed locally. In this context, the problems at the edge devices levels, like data storage and processing capabilities, are usually addressed by forwarding the data to the next computational level, benefiting from better hardware equipment [16]. At the same time, the applications can be offloaded toward the edge levels to increase responsiveness and address latency and bandwidth problems.

*2.1. Edge computing*

Edge computing puts computing and data storage near where they are used, usually at the network edge as opposed to cloud computing, where computing and data storage are done far away in distant data centers. Cao et al. [17] argue that all edge computing definitions focus on providing services and performing calculations at the network edge closer to the data generation source to meet the critical needs of industry and real-time applications. The edge devices such as IoT devices, smartphones, sensors, and other equipment that generate or consume data often have limited processing and storage capabilities, however, they form an edge infrastructure, which includes the hardware and software resources deployed at the edge.

The architecture of an edge computing network consists of the terminal layer, boundary layer, and cloud layer [17], [18]. End devices, such as sensors and actuators, are positioned at the terminal layer of the computing structure. This front-end environment offers increased interactivity and enhanced responsiveness for end users. Leveraging the available computing capabilities through the numerous nearby end devices, edge computing can deliver real-time services for certain applications. However, given the limited capabilities of these end devices, most demands cannot be met within the terminal layer. Consequently, in such instances, the end devices forward the resource requirements to the edge servers, located in the near-end (boundary) layer, where most of the data computation and storage migrates. The edge servers have better computing and storage capabilities, but they are also constrained compared to the cloud servers. This is why the computationally intensive tasks are forwarded to the cloud servers, deployed in the far-end (cloud) layer, but this can result in a significant latency penalty.

Edge devices feature a high degree of heterogeneity, leading to interoperability challenges, a significant obstacle in successful edge offloading. Additionally, network heterogeneity, caused by the diversity of communication technologies, affects edge service delivery. Consequently, ensuring that all edge devices and servers can work together seamlessly is crucial, pushing standardization and interoperability protocols importance for the edge computing ecosystem. The low latency and high bandwidth are the primary motivations for edge offloading to reduce the delay in sending data to a remote cloud server and receiving a response [18]. This is important for applications that require real-time or near-real-time processing, like some of the time-critical energy management services of the smart grid. Edge offloading aims to achieve bandwidth optimization by reducing the amount of data that needs to be transmitted to central data centers or the cloud [19]. Thus, edge servers aim at implementing relevant decision-making processes, based on which only already processed or relevant data, which can be particularly important in scenarios with limited network capacity, is sent to the cloud [8].

Finally, the additional tradeoffs need to be addressed, such as energy efficiency, security, and offloading overhead [20]. Given that many edge devices are battery-powered or have limited power resources, energy consumption is significantly lower in edge-based infrastructures than in cloud data centers. Furthermore, since data is processed closer to the source, there is potential for improved data privacy and reduced exposure to security threats [21].

## 2.2. Fog computing

Fog computing distributes services and resources of data processing, storage, and communication throughout the entire path from the cloud to the connected devices [22]. The main difference compared to edge computing is the hierarchical nature, offering a comprehensive range of computing, networking, storage, control, and services [23]. Consequently, fog jointly works with the cloud and edge nodes, representing the intermediate layer between the near-end and the far-end layers of the general edge architecture. A fog node includes multiple physical devices that offer resources and services and link the edge and cloud environments [24]. Fog nodes are responsible for processing, storing, and transmitting data supporting the offloading towards the network edge [25]. Fog nodes can be placed close to the data source to reduce the latency compared to traditional cloud computing or can be closer to the cloud to provide higher computing power and storage capabilities.

Fog nodes collaborate in a mesh fashion to offer load balancing, resilience, fault tolerance, data sharing, and reduced reliance on cloud communication [26]. Fog computing systems typically comprise three internal tiers but can include more tiers for specialized applications [22]. At the edge, fog nodes focus on data acquisition, normalization, and sensor and actuator control. In higher tiers, they handle data filtering, compression, and transformation, while nodes near the cloud aggregate data and generate further knowledge. Architecturally, edge fog nodes require less processing and storage but rely on substantial I/O accelerators for sensor data intake. With more tiers, each level extracts valuable data and executes more computationally intensive tasks.

Fog computing aims to establish a cohesive range of computing services extending seamlessly from the cloud to edge devices, as opposed to the base principle of edge computing which considers network edges as separate, isolated computing entities. Furthermore, fog provides stronger computing and storage resources than edge does. Thus, a fog node can aggregate data collected and processed by multiple edge nodes.

*2.3. Smart Grid*

The shift towards a renewable-based energy system impacts the electric power system operation that needs to integrate new ICT paradigms, models, architectures, and services to support decentralization [27]. The smart grid concept is fully connected to the dynamically interactive real-time infrastructure incorporating IoT and ICT-driven solutions everywhere, from electricity generation to delivery and consumption [28] (see Figure 2). Moreover, in decentralized scenarios digital communication and technology are mandatory to enhance the efficiency, reliability, and sustainability of electricity production, distribution, and consumption [29].

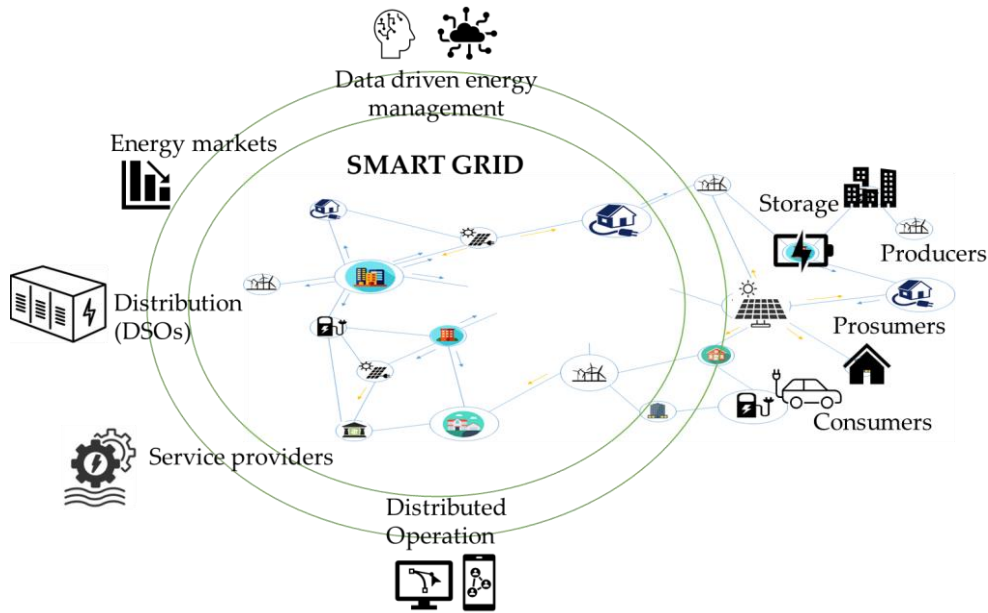

**Figure 2.** The smart grid ecosystem architecture

Smart grids aim to solve traditional grid problems caused by the energy transition, such as managing the uncertainty of renewable energy sources, demand management, shifting and shaving, congestion management, reducing power losses, and secure, efficient, and resilient services offering. Consequently, smart grid development has added smart data processing capabilities to the electrical grid [30] to improve the reliability and efficiency of the electric grid, optimize the grid operation and resources, integrate distributed RES, and deploy new technologies for improved metering and automation. On smart grids, data-driven services are implemented to enable real-time management during normal and emergency conditions [31]. They allow for grid decentralized operation within their safe ranges and reduce the overall costs with energy [32].

The management tools built on top of smart grids require the integration of advanced IoT devices, smart meters, data hubs, and storage systems, as well as AI-driven processes and decentralized components and architectures [28]. The deployment of smart metering devices and the renewable sources integration may increase the adoption of edge AI [33]. However, new challenges emerge in terms of data processing scalability and concerns about data privacy and security [24]. By connecting millions of devices, big data is fed into the distributed grid management systems, thus the tradeoffs related to latency, bandwidth, and response time need to be carefully considered.

## 3. Architectures Overview

Offloading concept in smart grids typically refers to the process of shifting computational tasks or data processing from local devices to remote servers or cloud platforms and back. In this section we start by analyzing the most relevant architecture for smart energy grids and then various computational architectures that have been proposed and can eventually facilitate the offloading of applications across computational continuum.

*3.1. Smart Grid Architectures*

There are two widely used smart grid architectural frameworks providing a structured approach for designing applications architectures, futures infrastructures, and reference scenarios: the European smart grid architecture model (SGAM) proposed by CEN-CENELEC [34] and the American smart grid conceptual model proposed by NIST (National Institute of Standards and Technology) [35].

SGAM is a layered model defining several interoperability layers [34] (see Figure 3). The asset and component layer models the energy assets and resources installed in the smart grid as well as the communication infrastructure for data exchange. The information layer defines the data flows and storage aspects of the infrastructure. The function layer addresses the functional capabilities needed to meet business objectives, while the business layer models processes, stakeholders, and objectives. The security layer spans across all layers offering features for security and privacy.

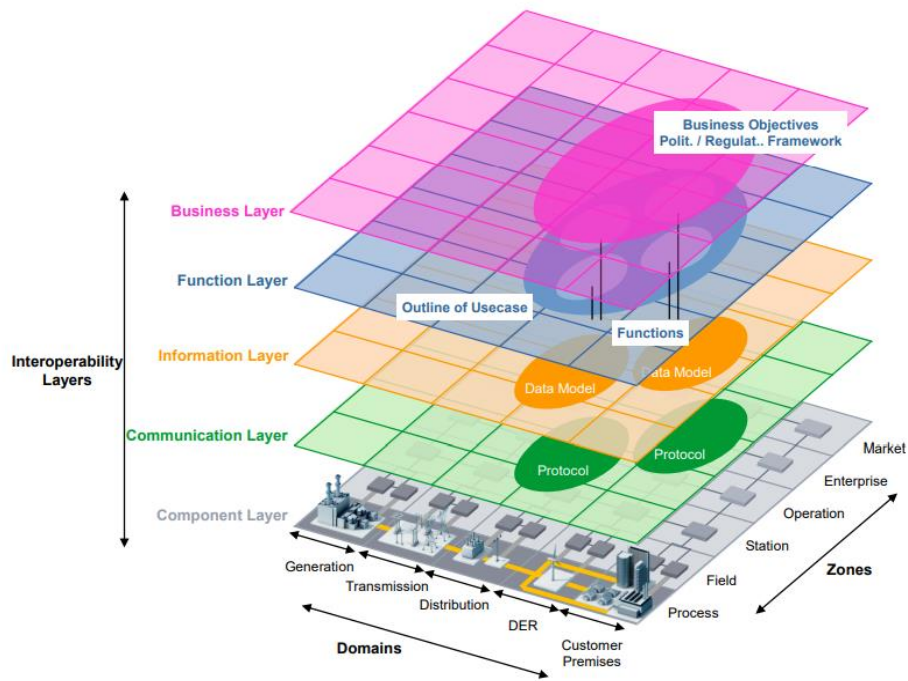

**Figure 3.** The SGAM architecture [36]

It considers different electrical grid domains such as generation, transmission, distribution, DER (distributed electrical resources), customer premises, etc. and foresees different aggregation zones. The data from the field devices and meters is usually aggregated or concentrated in the station zone to reduce the amount of data to be communicated and processed in the operational zone. At the same time the spatial aggregation can be done from distinct location to wider area for example multiple decentralized energy resources for a microgrid, smart meters in customer premises are aggregated in the neighborhood or community, etc. Being a reference architecture, SGAM offers several advantages for designing and implementing decentralized smart grid scenarios as it provides a common foundation, facilitates comparative analysis, and includes a specific mapping methodology [37-39].

The smart grid conceptual model proposed by NIST [35] offers a reference model to guide the development and interoperability of the smart grid, addressing aspects related to ICT models and architecture design and integration, paving the way for decentralized management scenarios (see Figure 4). The conceptual model explains the roles and services of smart grid in different domains and sub-domains that feature various services, interactions, and stakeholders who interact and communicate for achieving overall system objectives. Examples of such services are demand management, distributed generation aggregation, and outage control.

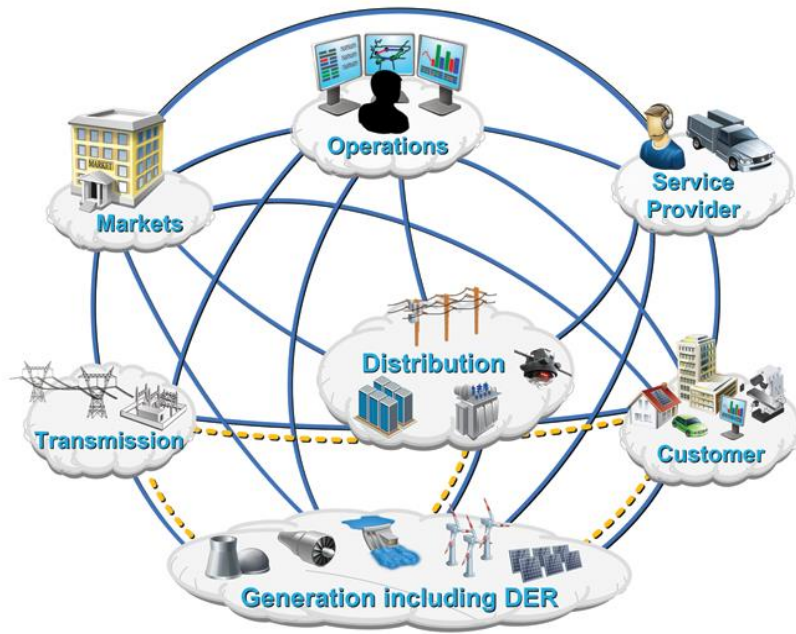

**Figure 4.** The NIST smart grid model [35]

The customer domain represents the end users of electricity, consumers, producers, or prosumers that can consume, generate, and store or manage energy. As classification the model considers three classes of customers, each having a different sub-domain: residential, commercial, and industrial. The customers boundaries are the smart meters and the energy services interface (ESI). The markets are economic mechanisms and facilitators that offer functionalities for actions to optimize system operation such as energy selling/buying, storage, etc. The markets domain allows to balance supply with demand within the smart grid and can use advanced Peer to Peer (P2P) trading mechanism based on modern technologies such as blockchain. The entities offering services to the involved actors are marked in the service provider domain. These business activities include usual utility services, like billing and customer accounts, and improved customer services, like controlling energy use, demand response and energy generation. Operations deal with the administrators of electricity movement such as smart grid managers and involve complex energy management systems to analyze and efficiently operate the grid; transmission refer to the carriers of electricity on long distances such as Transmission Systems Operators (TSOs) while distribution is the domain for distributors of electricity such as Distribution System Operators (DSOs). Generation (updated with DER inclusion in version 4.0 of the model) refers exclusively to energy producers including traditional generation sources and DERs. This domain includes all required technologies and infrastructures for generation/storage and participation to demand response programs.

The two architectural frameworks discussed above have been influential in developing smart grid architectures for different cases and scenarios including decentralization aspects [40]. Relevant architectures have been designed to address more in-depth different challenges related to computational continuum such as data management and distribution latency, and security [2], [3],[41]. However, it is worth noting that even though those architectures have benefits related to edge offloading, the implementation can be a complex process and might need a significant upfront investment for the initial setup [42], [43]. The integration of different smart grid layers with computational ones, such as edge or fog, and security management for unauthorized access are additional challenges that need to be considered [44]. Table 1 highlights important characteristics of relevant smart grid architectures.

**Table 1.** Characteristics of smart grid architectures

| Smart grid Architecture | Grid Automation | Grid Resilience | Protocols | Scalability | Fault Tolerance | Real-time Monitoring | Security Measures |
|---|---|---|---|---|---|---|---|
| Cloud-edge [41] | Yes | Yes | MQTT, IEC 61850 | High | Yes | Yes | Advanced Encryption |
| Three-Tier [2] | Yes | Yes | OPC UA, DNP3 | Moderate | Yes | No | SSL/TLS |
| Edge based with AI [3] | Yes | Yes | CoAP, IEC 60870 | High | Yes | Yes | Blockchain |

Mehmood et. al. defined a smart grid architecture consisting of four layers: the device layer, edge computing layer, cloud computing layer, and security layer, each layer representing a specific purpose for the data collection, preprocessing, storage, analysis, and security management [41]. The device layer consists of sensors, tags, actuators, and smart meters that collect data from the smart grid. The edge computing layer is located at the network edge with a primary goal of filtering and preprocessing data from the device layer before sending it to the cloud. The cloud computing layer is responsible for storage, computational analysis, and providing different application services while receiving only the summarized data sent from the edge nodes for global analysis. The security layer is responsible for the security in the smart grid, and it should be considered from the early stages of development, including network, computing, and memory management. The framework aims to address challenges such as data management, latency, security, and privacy for the smart grid system based on IoT, improving efficiency, reliability, and integration of renewable energy sources.

Feng et al. have proposed a three-tier architecture for the implementation of electrical engineering scenarios in smart grids [2]. The architecture foresees the Thing, Edge, and Cloud tiers. The Thing tier is responsible for the electrical equipment and communication access, executing specific operations in the smart grid, and implementing control orders. The Edge tier acts as an intermediary layer between the smart grid control center and the things, hosting resources for storage, communication, and computing. The energy resources are categorized into sub-layers based on their locations. While the low-power and fine-performance resources are positioned in the proximity of the things, resources with more robust computing capabilities are located closer to the control center. Additionally, the Cloud tier represents the Cloud computing resources that empower the computational and offers storage capabilities for the smart grid, offering monitoring solutions.

Molokomme et al. proposed an architecture involving multiple components such as residential, commercial, and industrial devices, edge servers, power systems, IoT devices and the overall cloud infrastructure [3]. The architecture integrates with Edge computing, introducing intelligence for analysis, monitoring, and processing data at the network's edge. The edge servers offload the tasks that require significant computation from devices with limited resources, improving the speed and processing capacity of the system. The architecture extends the features with AI algorithms deployed at the edge to improve communication, processing, and caching within the system. The objective is to raise awareness about security threats, manage power resources in an efficient way, and detect potential issues within smart grid systems. The architecture may utilize advanced optimization techniques such as federated learning, deep reinforcement learning, and peer-to-peer to enhance the performance and resource usage of the entire system.

*3.2. Edge Offloading Architectures*

The architecture addresses the design of systems that involve the distribution of computational tasks between different computing resources, such as edge devices, cloud servers, or other remote processing units.

Several architectures are based on combining different computing models, such as edge and cloud, to enable distributed computing and task offloading [19], [45], [46]. They focus on distributing tasks efficiently and optimizing resource usage. However, these benefits come with challenges such as increased dependency within the network and the management of tasks across multiple layers. Kaur et. al. proposed the KEIDS scheduler [45] for managing containers on edge-cloud nodes in the Industrial Internet of Things (IIoT) environment. The Edge nodes are responsible for collecting data from IIoT devices and performing initial processing. The Cloud Nodes empower the processing and storage capability for more complex tasks. The KEIDS controller acts as a central management and scheduling component, with the main objective of improving the allocation of tasks to the available nodes. The controller considers different factors such as carbon footprint, interference, and energy consumption in the decision-making process of scheduling. By optimizing energy utilization and minimizing interference, the scheduler aims to provide optimal performance to end-users in terms of application execution time and utilization. The architecture processes data in real-time and offers more flexibility and scalability in the ecosystem of edge-cloud for IIoT. Kovacevic et. al. [19] demonstrates the utilization of multi-access edge computing servers closer to mobile networks, transferring computation and storage from mobile and IoT devices. The edge servers are distributed across the radio access network and contain modest computational capabilities compared with cloud services. The offloading decision aims to minimize the usage of resources while concurrently maximizing the number of accepted requests that are time critical. The architecture stresses computing power and transmission with latency constraints for computation offloading requests. The objective is to optimize resource allocation to reduce the network traffic and service latency, while enhancing the resource utilization and acceptance rate. Nguyen et al. presents a resource adaptive proxy [46] in an edge computing environment consisting of multiple components, including the controller manager, scheduler, master API server, cloud controller manager and cloud edge client. The resource adaptive proxy component is implemented in each worker node of the Kubernetes (K8s) cluster and is integrated into every worker node within the cluster. The adaptive proxy algorithm consistently gathers resource availability, including CPU and RAM, along with network delays between edge nodes, to inform optimal load-balancing decisions. When making load-balancing decisions, the adaptive proxy considers the application resources available on each edge node. While local nodes are given priority for handling requests, in cases of local node overload, requests are directed to the most suitable edge node to minimize delay. The architecture is designed to reduce request latency and enhance overall throughput within the edge computing environment.

Several architectural designs leverage on a hierarchical distribution to achieve optimal task placement and enhanced QoS (Quality of Service) [47-50]. Pallewatta et. al. proposes a distributed architecture for IoT applications, utilizing microservices architecture and fog computing [47]. This framework facilitates the transition from monolithic application to distributed architecture for cloud deployment and task distribution to fog computing. It optimizes high-quality service delivery by strategic placement of microservices. Fog computing, in combination with resource-efficient deployment at the network edge, addresses the latency and bandwidth challenges of IoT applications. Moreover, the architecture allows for the dynamic composition of scalable microservices for achieving optimal performance in fog-based environments. Nevertheless, coordinating tasks across multiple layers can be a challenge due to the complexity of management. Firouzi et. al. proposed an edge layer design responsible for communication between sensors and nodes, as well as dedicated interconnections between fog nodes and the cloud [48]. The support for wireless connectivity in nodes relies on several factors like geographical location, data throughput, mobility, coverage, environmental conditions, spectrum licensing, and energy sources. The architectural viewpoint concerning control and management encompasses life cycle management, registration, provisioning, automated discovery, offloading, load balancing, task placement, task migration, and resource allocation. This hierarchical structure facilitates the dissemination of intelligence and computation, encompassing AI / ML, and big data analytics, to attain optimal solutions within specified constraints. This framework addresses challenges arising from the convergence of IoT and cloud computing, such as bandwidth limitations, latency issues, and connectivity concerns. Dupont et. al. [49] introduced the concept of IoT offloading, wherein containers are instantiated either at the edge or in the cloud, diverging from deployment on the gateway itself. The realization of this architectural model leverages OpenStack as a VM manager and Kubernetes as a container manager. Within the OpenStack environment, three Controller nodes and two Compute nodes are configured, with Kubernetes installed within the latter. The Kubernetes cluster encompasses Cloud nodes, Edge nodes, and two IoT Gateways as distinct nodes. These IoT Gateways are constructed using a Raspberry Pi version 3 along with an extension shield capable of supporting diverse wireless communication modules. The gateways deploy ARM versions of Docker and specific editions of IoT function containers tailored for both ARM and i386 architectures. The central orchestrator, utilizing Kubernetes labels, ensures the deployment of the correct container based on the target architecture. The discovery container initiates communication through Bluetooth Low Energy (BLE) hardware devices accessible on the gateway. An event notice is communicated to the orchestrator upon device detection, triggering subsequent processes. Taherizadeh et. al. proposed an architecture to optimize smart IoT applications, focusing on achieving elevated Quality of Service (QoS), flexibility, and dependability [50]. This framework introduces the concept of Microservices, where each business capability is encapsulated as a self-contained service with a clearly defined REST API. Employing lightweight container technologies like Docker, the architecture virtualizes and implements the necessary Microservices. Components include a Container Orchestrator, an Edge-Fog-Cloud Monitoring System, and infrastructure elements. This Edge-Fog-Cloud architecture ensures that data processing and computation occur at the most suitable level, enhancing performance, reducing latency, and elevating Quality of Service (QoS) for IoT applications. The framework facilitates the orchestration of Microservices, seamlessly transitioning from Edge computing nodes to Fog and Cloud servers within the geographical vicinity of mobile IoT devices. In comparison to fixed centralized Cloud providers, this distributed computing architecture delivers swifter service response times and enhanced QoS.

Relevant edge orchestration architectures have a primary focus on the organization and scheduling of tasks across both cloud and edge nodes [51], [52]. These approaches offer advantages such as resource optimization, but the creation of efficient orchestration strategies can be a complex task. Böhm et. al. defines an architecture based on a container registry that contains the images of applications and is used to design the nodes within the cloud infrastructure [51]. The autonomic controller distributes responsibilities to various nodes. The distribution is based on a defined strategy, algorithm, or policy. To distribute the applications across both cloud and edge layers, diverse provisioning models are used. Orchestration across both cloud and edge layers ensures a strategic distribution of applications in edge, cloud, and IoT components. The distribution is based on a set of objectives, adopting a multi-objective approach to support optimal efficiency. To offer this framework, complex optimization and scheduling models are required, with the capability of dynamically allocating applications based on resource demand and supply. Pérez et. al. defines intelligent container schedulers for different interfaces within cloud-fog-IoT networks [52]. The schema consists of three primary interfaces: cloud-to-fog, fog-to-IoT, and cloud-to-IoT, each with distinct responsibilities and functionalities. It emphasizes the importance of designing and implementing microservice schedulers for these interfaces, offering several benefits, including the optimization of runtime, adherence to latency restrictions, power consumption reduction, and load balancing. The schema visually demonstrates the complexity of the network architecture and the need for tailored scheduling strategies for each interface.

Finally, serverless edge computing architecture [53] emphasizes the integration of Serverless functionality to manage event processing, with the goal of reducing the need for extensive adjustments for IoT devices. However, it is worth mentioning that Serverless functions may encounter delays when starting up, which can affect their ability to respond promptly when initially called upon. Moreover, the restricted duration of execution for Serverless functions may function as a limitation for specific applications. From the edge perspective, IoT devices connect to edge nodes with Serverless functionality for efficient event processing while IoT devices require minimal adaptation, following function-based principles. In the cloud, Serverless integrates with its edge counterpart. Table 2 presents an overview of relevant computational offloading architectures, and their characteristics.

**Table 2.** Characteristics of Offloading Architectures

| Category | Architecture | Application Orchestration | Technology | Serverless Computing | Dynamic Offloading | Multi-layer Coordination | Cost-Efficient Scaling | Optimization Algorithms |
|---|---|---|---|---|---|---|---|---|
| Edge-Cloud Integration | KEIDS [45] | Yes | Docker, Kubernetes | No | Yes | Yes | Yes | Yes |
| | Multi-access Edge Computing [19] | Yes | Docker, Kubernetes | No | Yes | Yes | Yes | No |
| | RAP [46] | Yes | Docker, Kubernetes | No | Yes | Yes | Yes | Yes |
| Edge-Fog-Cloud Integration | Hierarchical Edge-Fog-Cloud [48] | Yes | Docker, Kubernetes | Yes | Yes | Yes | Yes | Yes |
| | IoT Offloading [49] | Yes | Docker, Kubernetes | No | Yes | Yes | Yes | No |
| | Edge-Fog-Cloud for IoT [50] | Yes | Docker, Kubernetes | Yes | Yes | Yes | Yes | Yes |
| | Fog Computing with Microservices [47] | Yes | Docker, Kubernetes, KubeEdge | No | Yes | Yes | Yes | No |
| Edge Orchestration | Autonomic Controller [51] | Yes | Docker, Kubernetes | Yes | Yes | Yes | Yes | No |

| | Intelligent Container Schedulers [52] | Yes | Docker, Kubernetes | Yes | Yes | Yes | Yes | Yes |
| Serverless Integration | Serverless Edge Computing [53] | Yes | AWS Lambda, Azure Functions | Yes | Yes | Yes | Yes | No |

## 4. Offloading criteria in smart grid

In this chapter, we explore key factors and variables that serve as guiding principles in making informed decisions to determine whether a task or process should be offloaded from a cloud environment to a local edge device or on-premises system. When discussing offloading in the context of smart grids, the decision-making process is crucial for optimizing resource utilization, improving performance, and minimizing costs across the computing continuum. The process is affected by several variables or factors such as network performance, data and AI processes, computational requirements, application-specific factors, and energy efficiency.

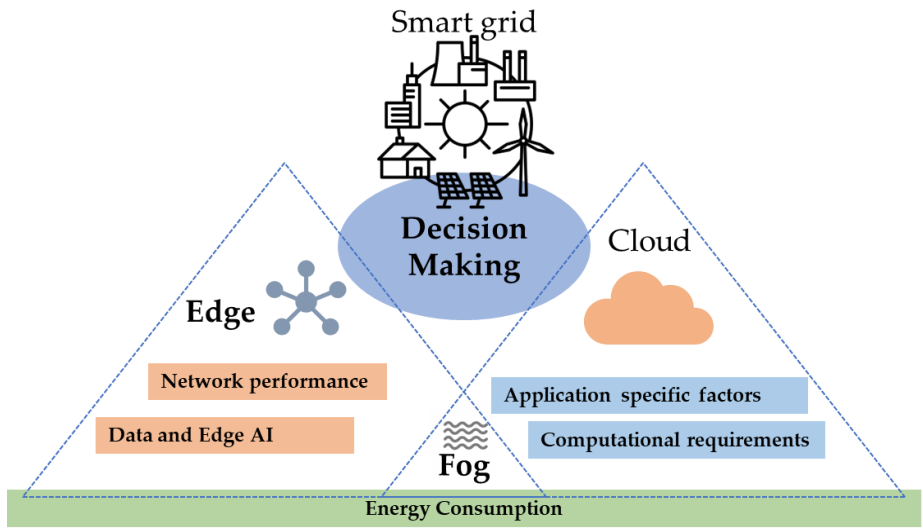

**Figure 5.** Offloading decision making in smart grid

*4.1. Network Performance*

Latency, available bandwidth, and response time of a distributed task are important performance metrics in the context of edge offloading for the energy sector [54]. These metrics help to identify the efficiency and effectiveness of making edge offloading decisions and to assess the performance of applications. In the current literature, several papers proposed solutions applicable to smart grid scenarios addressing aspects such as the high latency that may limit the ability to react in smart grid real-time control of assets [55-57].

Wang et al. [55] proposed a holistic approach to assess the requirements of different energy services in smart grids. Metrics such as latency, bandwidth, and response time, are used to create schemes for allocating resources and establishing priorities. They help in making offloading decisions and reducing costs related to task execution delays. The results highlight the efficiency of offloading strategies based on multi-attribute preferences, emphasizing how performance metrics enhance business outcomes in smart power services. Smart grid components and their performance models influence task execution time, showing the importance of considering energy and latency trade-offs [56]. Network performance metrics quantitative measures can be used to optimize the smart grid performance and energy efficiency. Their assessment can be used to determine optimal offloading strategies based on specific smart grid requirements, considering factors like energy consumption, response time, and availability [57].

The fog computing infrastructures are used as an intelligent gateway within an IoT framework and offer an effective solution to reduce the latency of applications in edge computing [7]. A multi-period deep deterministic policy gradient algorithm can find an optimal offloading policy to reduce computation, transmission delay, and energy consumption for a collaborative cloud network [58]. Longer communication latency can lead to delays in data transfer, impacting the response of time-critical applications in smart grid. Markos et al. investigate different communication strategies for edge offloading and their impact on energy use and response time [59], concentrating on offloading decisions of the computational tasks for a mobile cloud environment. A multi-objective service provisioning scheme is defined to enhance the overall performance of both network and computation infrastructure while maximizing the usage of the battery lifespan of mobile devices.

Huaming et al. [57] propose the Energy-Response time Weighted Sum and Energy-Response time Product metrics to provide a balanced approach while assessing the tradeoff between energy consumption and response time. The metrics combine additive and product factors, prioritizing both aspects without being influenced by different operational scales. Kovacevic et al. [19] emphasize the relevance of performance metrics for assessing applications with critical delays and decision-making related to offloading to improve collaborative resource sharing among cloudlets and mobile cloud providers. Decision-makers can use metrics such as latency, resource utilization, acceptance rate, and resource sharing for efficient cloud offloading in smart grids. Jyothi et al. [60] proposed a dynamic programming solution to offloading using the Hamming Distance Termination. They showcase a strategy for efficiently offloading specific tasks to the cloud, thereby improving execution time, and optimizing energy usage. Bandwidth is crucial for proper utilization and efficient data transfer in the cloud. Insufficient bandwidth can lead to performance issues and hinder the overall system's performance [61].

Huaming et al. [62] use Lyapunov optimization to minimize energy consumption while ensuring that response time meets a given constraint. The prolonged latency of cloud offloading is not suitable for real-time requirements while direct edge offloading relies on powerful edge servers, which may not be practical for prosumer households in smart grid scenarios [63]. User-centric perspectives and quality-of-experience-based cost functions have also been considered to optimize the energy-latency trade-off [64]. The shift towards cloud computing has led to the definition of architectures susceptible to latency at different levels [65]. A delayed offloading model has been devised to harness the capabilities of Wi-Fi and mobile networks, considering energy efficiency, performance metrics, and intermittently available access links [57]. Finally, various offloading techniques are defined by Akram et al. [66], including round robin, odds algorithm, and ant colony optimization. These techniques can enhance overall smart grid system performance, addressing network performance, reliability, stability, and energy efficiency.

*4.2. Data and Edge AI*

Cloud offloading in the energy sector involves considering the location (physical or logical) and data characteristics such as volume, velocity, and variety. Decisions regarding offloading can be determined by considering the contextual aspects of the data, encompassing requirements like low response time and various other performance characteristics [7]. The renewable energy sources adoption at the edge of the grid and the integration of IoT sensors and actuators in smart grids require real-time processing [2] and the definition of new AI and data-driven energy services. Edge AI is emerging as a new paradigm for the efficient management of smart grids due to machine and deep learning model improvements [8]. Also, it is facilitated by the recent advancements in computing infrastructure towards edge data processing and the adoption of IoT devices in the smart grid that generate big data that need to be processed and considered by AI [15].

The distribution of AI models towards the edge aims at reducing latency and integrating with AI-driven management services. In some scenarios, location information can be post-processed in the cloud using raw GPS signal data, resulting in lower energy consumption for location tagging [67]. The objective of computational offloading for traffic in mobile cloud computing is to enhance the performance of both network and computational infrastructure, all while adhering to latency constraints [59]. Offloading strategies within mobile cloud computing strive to optimize effectiveness by transferring workloads either to adjacent cloudlets or to distant cloud computing resources [68]. Energy-aware offloading protocols and architectures are being explored to cope with the increasing number of mobile applications and limitations of battery technologies, with a focus on cloud resource management and green computing [69].

Federated frameworks offer promising models for processing data using ML algorithms at the edge nodes and orchestrating a global model in the cloud [11]. However, their applications in smart grid scenarios and integration with new real-time context-aware energy asset management services are limited. Computation offloading frameworks can meet the performance requirements of IoT-enabled services by considering context-based offloading [70]. The offloading decision should consider contextual information to improve accuracy and performance [71]. The dynamic nature of the edge mobile computing environment poses challenges, but a context-sensitive offloading system using machine learning reasoning techniques can provide accurate offloading decisions [72]. In adaptive offloading systems, energy optimization can be achieved by including context-specific optimization on mobile devices and offloading computational components to a high-performance remote server or the cloud [73]. Current research is focused on improving offloading protocols and architecture to be more energy and contextual-aware. It also enhances scheduling and balancing algorithms to achieve intelligent solutions in the context of edge cloud offloading in the energy sector [74].

*4.3. Computational Requirements*

In IoT-based applications, the node processing capabilities can influence the decision to offload specific tasks from the cloud to the edge and back. These factors are crucial for determining the feasibility and efficiency of the offloaded tasks from edge devices to the cloud infrastructure. The selection of the processing node across the computing continuum depends on several factors: CPU information, memory information, network state information, and average network delays [75]. The challenge is to take optimal edge computational orchestration decisions under uncertain and dynamic conditions [76] impacted by the need for resources in terms of bandwidth and memory, potential failures such as data network issues, network speed, energy consumption, and the lifespan of applications.

The decision to offload specific models is made based on the characteristics and execution patterns of tasks, considering the limitations of the resources for the edge devices and the communication cost between the device and the cloud [77]. The offloading decision algorithm can integrate multiple parameters to reduce application response time, reduce energy consumption, and extend the battery lifetime for the devices [78]. The offloading decisions can be improved by setting threshold values for processing time and employing adaptive algorithms that can dynamically adjust and ascertain the optimal threshold value, ensuring a balanced load on resource-limited devices and edge nodes. Limited resources and energy of edge devices require delegated tasks to the fog and cloud. The presence of augmented computing power and extensive storage capacity can manage the workload in a more effective way.

Automation is essential in managing cloud edge platforms in smart grids requiring applications virtualization, semantic integration, and data connectivity. Comprehensive orchestration techniques are needed to coordinate, schedule, and run applications across the edge-to-cloud network [4], [51]. This will help to deliver real-time energy services at the edge of the smart grid. The dynamic nature of resources in IoT computing farms needs a more robust control mechanism to ensure efficient operation [79]. The offloading architecture aims to minimize the delay while considering energy consumption limitations, and algorithms have been suggested to optimize the delay [80]. The performance improvement can be achieved by offloading computation in cloud robotics due to various factors such as parallel processing capabilities, the availability of resources in the cloud, and communication delays [81]. Based on these factors, decisions regarding offloading are influenced by an assessment of energy consumption, task processing power requirements, and the balance between local execution and offloading tasks.

*4.4. Application-specific Factors*

The impact of application type and application migration overhead on the cloud offloading decision is significant. However, the offloading decision process can introduce overhead when implemented on the mobile device. Shifting the offloading decision process to the cloud can reduce this overhead and improve energy savings and execution time [82]. Moreover, a decision-making system that considers the client's hardware and software resources, location, context, and security capabilities can support dynamic migration and improve the offloading procedure [83]. Different types of applications require cloud resources and services [84]. Selection of the most suitable cloudlet for offloading an application is crucial for reducing energy consumption and latency in application execution [85]. The migration of applications to the cloud introduces overhead and adaptation needs at each layer [86]. The challenges and solutions for migrating different parts of the application to the cloud should be considered, including considerations that apply across various aspects and possible tradeoffs before migration in a new environment [87].

In the context of smart grids, cloud offloading decisions are influenced by the application type and associated migration overhead, optimizing resource utilization, reducing costs, and meeting service level agreements [88]. Atta et al. [84] emphasize the importance of application type in determining the most suitable cloudlet for offloading tasks. Cloudlets demonstrate efficiency in processing diverse application types, impacting load-balancing demands and requiring distinct algorithms [89, 90]. To facilitate offloading based on application type, an approach for strategic cloudlet selection is introduced [91], aiming to minimize mobile terminal consumption and latency. This strategy also assists in load balancing by distributing processing tasks across multiple cloudlets, preventing overload on a single cloudlet.

In smart grids, pivotal roles are played by cloud migration technologies [92]. These technologies strategically place applications across geographically distributed cloud data centers, aiming to reduce costs and adhere to service-level agreements. The importance of considering application migration overhead, including factors such as execution time and energy consumption, is crucial in making informed offloading decisions [93]. Huijun et al. [94] monitor the performance of streaming applications, automatically adjusting the flow of the application graph by offloading computationally intensive operators to virtual machines in the cloud. The primary objectives are to optimize resource utilization and enhance the efficiency of smart grid applications. The research underscores considerations in cloud offloading decisions for smart grids. Finally, Seyedeh et al. [95] address problems related to application migration and service discontinuity to reduce application delay in hybrid cloud-fog systems. Additionally, factors such as application types, cloudlet selection strategies, migration overhead, and dynamic performance monitoring contribute to the intelligent optimization of smart grid operations, ensuring efficient resource utilization and overall system efficiency improvements.

*4.5. Energy Consumption*

Energy consumption plays a pivotal role in making informed decisions to offload specific tasks. Offloading is shifting computation from mobile devices to remote cloud servers, which can help enhance efficiency and minimize battery consumption [96]. The distributed energy-efficient computational offloading reduces data transmission size and energy consumption cost [70]. In fog computing, dual-energy sources, such as solar power and grid power, can support fog nodes and reduce the carbon footprint in IoT systems [97]. Pramod et al. [98] measure the file size and execution time to decide whether to execute the file locally or send it to the core cloud, considering both time and energy savings. Cloud-based software architectures are also being studied to achieve energy-efficient solutions, considering the complexity and investments required for migration and maintenance [99]. Overall, energy efficiency and power consumption play a crucial role in determining the most suitable offloading strategy [100]. Gu et al. [101] propose techniques for energy-efficient computation offloading in the context of 5G networks. Others use energy-efficient frameworks for cloud architectures, which can save up to 25% of the electrical consumption of cloud nodes [102].

Literature reviews on energy-efficient software architectures within cloud environments highlighted the crucial role of energy efficiency in provisioning cloud services [99, 103]. Han et al. [104] discuss the definition, principles, and challenges of implementing high energy efficiency in cloud environments. Mobile cloud computing empowers mobile devices to transfer their workloads to distant cloud servers, leveraging the abundant resources of the cloud to optimize efficiency [105]. Fog computing solutions are proposed to alleviate cloud computing's constraints in terms of latency and high bandwidth requirements by bringing resources closer to users [106]. Power management plays a crucial role in achieving power savings, and changes in architecture, topology, average load/server, and scheduling algorithms can significantly improve energy efficiency [107]. Table 3 shows a comparative analysis of the factors that should be considered when making offloading decisions.

**Table 3.** Decision variables in offloading.

| Decision Variable | Aspects | Approaches | Impact on Cloud Offloading |
|---|---|---|---|
| Network performance | Latency, Bandwidth, Response Time | FC in IoT architecture [7], MP-DDPG algorithm [58], Communication strategies and delayed offloading [57], multi-objective service provisioning [59], DPH algorithm [60], Lyapunov optimization [62], User-centric QoE [64] | Efficient offloading decisions based on minimizing computation, transmission delay, and energy consumption. Optimizing network and computation infrastructure while maximizing battery lifetime [59]. |
| Location and Data Characteristics | Location, Data volume, velocity, variety | Context-based offloading [70], CSOS with ML [72], Energy-aware protocols [73], Adaptive offloading [74], EMCO, MobiCOP-IoT, Autonomic Management [70], Contextual information utilization [71], Green computing [74] | Context-specific optimization by considering the context of data, reducing energy consumption, and achieving accurate offloading decisions [74]. |
| Computational Requirements | CPU, Memory, HDD, Devices, Processing Capabilities | Processing node selection [75], Models based on task nature [77], C-RAN architecture, Adaptive algorithms, Edge devices offloading [78], Control mechanisms [79], Offloading architecture [80], Computation offloading in cloud robotics [81] | Influencing feasibility and efficiency of offloading tasks, dynamic adjustment, and trade-off between local execution and cloud offloading [80]. |
| Application-specific Factors | Application Type, Migration Overhead | Decision support system [83], Cloud resources for different app types [84], Suitable cloudlet selection [85], Challenges in | Significant impact based on computational, storage, and bandwidth requirements. Overhead considerations for application |

| | | migration [86], Overhead and adaptation needs [87] | migration and service selection [87]. |
|---|---|---|---|
| Energy Consumption | Power Consumption, Energy Efficiency | EECOF [70], Dual-energy sources in fog computing [97], Measuring file size and execution time [98], Energy-efficient architectures [99], MEC energy-efficient computation offloading [101], Energy-efficient framework [102], Power management [107] | Crucial role in determining the offloading strategy based on distributed frameworks, file size, execution time, and efficient architectures [48]. Power savings through various approaches and technologies [107]. |

## 5. Decision Making Techniques

Several decision-making techniques can be employed in the context of cloud offloading to determine when and what to offload. The choice of technique depends on factors such as the application's characteristics, the dynamic nature of workloads, and the specific goals of offloading. We have classified the techniques based on their type in heuristic optimization-based ones and reinforcement learning based ones.

*5.1. Metaheuristic Optimization*

Metaheuristic optimization algorithms are a class of algorithms designed to find approximate solutions to optimization problems. These algorithms are often used when the search space is large, complex, and may contain multiple local optima. Metaheuristic algorithms do not guarantee an optimal solution but are effective in exploring solution spaces such as cloud offloading and finding good solutions in a reasonable amount of time.

Materwala et al. [108] optimize energy by redirecting requests from vehicles to both edge and cloud servers, thereby reducing the energy consumption associated with the vehicles themselves. It employs an evolutionary genetic algorithm to optimize the energy of edge-cloud integrated computing platforms. An adaptive penalty function is utilized by the algorithm to integrate optimization constraints into the genetic algorithm, ensuring that the offloading process meets Service Level Agreements. It refines its selection process by employing an adaptive fitness function that assesses the proximity of each solution to the optimal solution. The algorithm includes stages such as initialization of offloading solutions, evaluation of solutions, selection of fittest solutions, crossover to produce offspring solutions, mutation of server allocations, and termination. Through comparative analysis, the proposed algorithm demonstrates significant energy savings compared to random and no-offloading approaches, with a violation rate of only 0.3%. An algorithm for collaborative offloading among cloud, edge, and terminal devices, incorporating enhancements to a genetic algorithm, is introduced in [109]. The algorithm is structured as a non-linear problem in combinatorial optimization, striving to reduce the overall task consumption while ensuring compliance with computational delay constraints. The algorithm's system model encompasses various computational tasks, diverse mobile devices, multiple small-cell base stations, numerous micro base stations, and a cloud server. The proposed algorithm undergoes theoretical analysis, verification, and comparison with other algorithms using simulation trials. The results indicate superior performance, particularly when considering diverse quantities and capabilities of mobile devices and servers at the network edge. The iNSGA-II tasks offloading mechanism in [110] adopts a metaheuristic-based approach, utilizing the non-dominated sorting genetic algorithm (NSGA-II) within edge/cloud networks for serving mobile applications. Addressing the task offloading challenge as an NP-hard problem, the mechanism centers on relocating computationally intensive tasks from mobile devices to edge servers. Enhancements to the crossover and mutation operators facilitate faster convergence, setting it apart from other evolutionary algorithms. Employing NSGA-II as a population-based metaheuristic, the mechanism efficiently determines task-offloading decisions within a reasonable timeframe. Numerical outcomes under simulated workloads underscore the cost-effectiveness of the proposed mechanism, enhancing the average utilization of edge servers and reducing energy consumption and execution time compared to alternative task offloading approaches based on metaheuristics.

The SA-BPSO algorithm presented in [111] breaks down the optimization problem into three distinct sub-problems: the allocation of computing resources, the allocation of uplink power, and task offloading. Convex optimization techniques are employed to optimize computing resource allocation, while the bisection method is applied for uplink power allocation. The SA-BPSO algorithm encodes and initializes the particle swarm, maps the velocity to the interval [0,1] using the Sigmoid function, and binary encodes the position of each particle. The SA-BPSO algorithm effectively reduces the total user overhead compared to other schemes and ensures user quality of service. An innovative approach called E-PSO) algorithm designed in [112] for optimization of energy consumption of virtual machines in cloud environments. The primary objective is to minimize energy consumption using the strategic placement of virtual machines in a specific location closer to data sources. It introduces a locally aware fitness function focused on energy considerations and formulates a coding scheme for relocating virtual machines. The proposed E-PSO algorithm identifies an optimal VM replacement strategy aiming to reduce energy consumption. The E-PSO algorithm achieved a 22% reduction in overall energy consumption.

A recursive version of ant colony algorithm called RACO introduced in [113], with the primary objective to address issues related to energy consumption and potential service-level agreement violations in the context of cloud computing. The RACO algorithm consists of monitoring and refreshing pheromone levels, selecting cities, and guiding the most optimal ant's movement. In the monitoring pheromone step, the algorithm keeps track of the pheromone levels and updates the ant's movement toward the optimal solution. The city selection step involves the ant selecting the next city to move based on the pheromone levels and the distance between cities. The guidance of the most optimal ant includes the selection of the best ant, which moves randomly through the cities to discover the optimal solution. The gap between movements of the best ant aims to mitigate EC and SLA violations. The outcomes indicate a substantial reduction of approximately 40-42% in EC when comparing RACO to the conventional ACO algorithm on Planet Lab. Efficient Ant Colony Cloud Offloading Algorithm (EACO) is developed in [114] to reduce energy consumption while considering completion time constraints. The algorithm divides mobile applications into fine-grained tasks with sequential and parallel topology. It focuses on tasks scheduling between execution on the mobile device and offloading to the cloud to limit the increase in completion time. EACO achieves an average energy reduction of 24%-59% compared to previous work, with a corresponding increase in completion time of 3.6%-28%. In [115], the ant colony algorithm is proposed for optimizing energy usage when allocating resources to virtual machines (VMs). The algorithm effectively reduces energy consumption and minimizes environmental impact. The use of pheromones by the ants guides their decision-making process as they deposit them along their paths. The algorithm experiences iterative updates of pheromone levels until the quality of solutions discovered by the participating ants improves. The ant colony algorithm achieved an average energy reduction of 24%-59% compared with other works.

Samoilenko et al. [116] introduce the whale optimization approach to address challenges in task offloading within a cloud-fog ecosystem. This method involves dynamic offloading decisions made at runtime, utilizing the whale optimization algorithm to enhance various quality of service metrics, such as delay and energy consumption. It employs a population of solutions, represented as Whales, to find the best solution to a given problem. The exploitation whale optimization algorithm defined in [117] is an effective optimization algorithm that combines the differential evaluation and whale optimization algorithms to find optimal solution. By combining the exploration capabilities of whale optimization and the exploitation capabilities of differential equations, the algorithm solves the limitations of conventional heuristic algorithms, such as uncertain convergence time, lower exploration and exploitation ability, and implementation difficulties. The spiral bubble-net hunting behavior observed in humpback whales helps the algorithm to identify the optimization strategy. The algorithm reduced both energy consumption and response time during the offloading process.

Yuan et al. [118] define a hybrid metaheuristic algorithm used for the concurrent optimization of computation offloading and resource allocation in mobile edge computing with the primary goal of reducing the total energy consumption. This optimization considers offloading ratio, CPU speeds, allocated bandwidth, and transmission power. The algorithm sets random numbers for particle positions and velocities within the particle swarm optimization framework. The fitness values of particles are updated using a penalty function method to convert constraints into penalties. Additionally, several operations perform mutation and selection using principles from genetic algorithms. Metropolis acceptance rule in Simulated Annealing (SA) updates the velocity and position. Fitness values recalculate the local and global optimal positions. The algorithm continues iterating until it fulfills a predefined stopping criterion. It requires reaching the maximum allowable iterations or having a specified percentage of particles attain uniform fitness values. The final solution converts the globally optimal position into decision variables.

*5.2. Reinforcement Learning*

The DDPG (Deep Deterministic Policy Gradient) algorithm, as described in reference [119], represents a model-free, off-policy reinforcement learning approach that emphasizes the advantages of deep neural networks and deterministic policy gradients. The algorithm combines computation offloading, service caching, and resource allocation to reduce energy consumption for a collaborative MEC system. It includes a Mixed-Integer Non-Linear Programming (MINLP) framework. The DPPG algorithm uses a deep neural network to identify the optimal policy of each decision variable. The critic network assesses the quality of the chosen actions, while the actor network selects the action (i.e., resource allocation, service caching, and offloading decisions) based on the current state. The critic network performs training to approximate the value function, representing the anticipated long-term reward for a specific combination of a state and action. The actor network performs training to maximize the foreseen long-term reward. The training process involves iteratively updating the actor and critic networks using the DDPG algorithm's update rules, which involve gradient descent and target network updates to stabilize the learning process. During the iterative training process, the DDPG algorithm acquires optimal approaches for offloading computations, caching services, and allocating resources. These findings reduce long-term energy consumption in the collaborative Mobile Edge Computing (MEC) system.

Weichao et al. [120] propose an adaptive strategy for task distribution addresses the challenge of environment adaptation. It optimizes objectives related to task latency and device energy consumption. The meta-reinforcement learning algorithm continuously explores and adjusts the edge environment. A task management network, structured on Seq2Seq neural network architecture, is constructed to handle diverse facets of task sequences. Introducing a first-order approximation method accelerates the computation of meta-strategy training for the Seq2Seq network. The algorithm reduces task processing delay and device energy consumption while adapting to needs. Results illustrate the algorithm's performance over existing methods across various tasks and network landscapes. Antoine et al. [121] defines the deep reinforcement learning algorithm for task offloading to solve the problem of computation offloading with task dependency represented as a directed acyclic graph within the collaborative scenario involving cloud, edge, and end systems, including multi-user environments, multi-core edge servers, and a dedicated cloud server. The Markov Decision Process supports the task offloading while Deep Reinforcement Learning incorporates action masking based on task priority. This algorithm uses the computational capabilities of both cloud and edge servers to derive optimal policies for computation offloading. It improves the average energy consumption and time delay experienced by IoT devices. Xin et al. [122] formulates the optimization challenge of collaborative computation offloading between the cloud and edge as a dynamic problem represented using a Markov decision process. It concurrently refines average delay, energy efficiency, and the revenue per unit time metrics and combines exploration and exploitation to identify the offloading strategy. The simulation results show the efficiency of the proposed algorithm, especially as the number of tasks offloaded for computation increases. Jie et al. [123] consists of decomposition of the (offline) value iteration and (online) reinforcement learning, which allows for learning in a batch manner and improves learning convergence speed and run-time performance. The algorithm learns the optimal policy of dynamic workload offloading and edge server provisioning to minimize the long-term system cost, including service delay and operational cost. It uses a post-decision state based learning approach, exploiting the structure of state transitions in the energy harvesting of the edge cloud system. Also, the algorithm enables the edge system to determine the optimal offloading and autoscaling policies and solves the "curse of dimensionality" problem associated with large state spaces in Markov Decision Processes. The simulation demonstrates significant improvements in how edge computing performs compared to fixed or short-term optimization methods and traditional reinforcement learning algorithms. Mashael et al. [124] utilizes a set of deep neural networks in a distributed manner to find near-optimal computational offloading decisions, aiming to reduce overall energy consumption in cloud offloading scenarios. The algorithm treats the problem as a binary optimization task. Due to the computational complexity of solving this NP-hard problem, an equivalent reinforcement learning form is generated. The distributed deep learning algorithm leverages parallel deep neural networks to find the near-optimal offloading decisions. Results from simulations illustrate that the suggested algorithm rapidly reaches convergence and significantly lowers the system's total consumption when contrasted with established benchmark solutions. Yongsheng et al. [125] introduce an offloading algorithm based on a deep learning network to calculate the most efficient offloading strategy, considering energy and performance constraints. The algorithm formulates energy and performance considerations into a cost function, and a deep learning network is trained to determine the optimal solution for the offloading scheme. It identifies the best set of components to offload to a nearby server, enhancing the computational capabilities of user equipment for running resource-intensive applications. Important findings illustrate the superior performance of the proposed approach compared to existing methods concerning energy and performance constraints. Sellami et al. [126] introduces a combination of blockchain technology with deep reinforcement learning. The main objective is to raise awareness of energy operations in a classical IoT framework using software defined networking. Utilizing policies, the approach optimizes various aspects while enhancing reliability, reducing latency, and optimizing energy efficiency. The proposed method prioritizes consumable energy and elevates Quality of Service in operations. Experimental results showing the improvements of network latency and energy efficiency compared with traditional algorithms.

The Deep Reinforcement Learning algorithm minimizes power consumption by making informed decisions during each time slot based on content request details and current network conditions in [127]. Addressing the issue as a power minimization model allows aggregation of requests and extensive in-network caching deployment. Leveraging past slot data and the present network state, the reinforcement learning algorithm enhances power efficiency in cloud-edge-end collaboration networks. Results highlight the performance of the proposed content task offloading model in power efficiency compared to current alternatives, demonstrating rapid convergence to a stable state. Table 4 presents a comparative analysis of main decision-making techniques used for cloud offloading.

**Table 4.** Decision making alternatives.

| Type | Algorithm | Approach | Optimization Target | Performance Metrics | Efficiency Improvement |
|---|---|---|---|---|---|
| **Metaheuristics** | Genetic Algorithms [108] | Offloading from vehicles to servers | Energy consumption, SLA compliance | Energy savings, low violation rate | Minimize energy consumption, meet SLAs |
| | Genetic Algorithm (IGA) [109] | Cloud-edge-terminal collaboration offloading | Task consumption, delay constraints | Superior performance, task completion within constraints | Minimize overall task consumption, meet delay constraints |
| | NSGA-II [110] | Task offloading in edge/cloud networks | Task offloading decisions | Faster convergence, cost-effective solution, energy reduction | Cost-effective task offloading, reduce energy consumption |
| | SA-BPSO [111] | Task offloading, resource allocation, power allocation | Total user overhead | Effective reduction in total user overhead, ensure QoS | Optimize task offloading, resource allocation, and power allocation |
| | E-PSO [112] | Energy-efficient VM consolidation in cloud | Energy consumption | Reduction of 22% in energy consumption | Minimize energy consumption |
| | Recursive ACO (RACO) [113] | Cloud computing energy reduction | Energy consumption, SLA violations | Reduction of EC by 40-42% compared to traditional ACO | Minimize EC and SLA violations |
| | Efficient ACO (EACO) [114] | Cloud offloading with completion time constraints | Energy consumption, completion time | Average energy reduction of 24%-59%, limited increase in completion time | Reduce energy consumption, limit completion time increase |
| | ACO for VM Allocation [115] | VM allocation for energy optimization | Energy consumption | Average reduction of 24%-59% in energy consumption compared to previous work | Minimize energy consumption |
| | Whale Optimization [116] | Task offloading in cloud-fog environment | QoS metrics (delay, energy consumption) | Improved QoS metrics, mimics social behavior of humpback whales | Improve QoS metrics, make runtime offloading decisions |
| | Exploitation WOA (EWOA) [117] | Offloading in mobile ad hoc cloud environment | Energy consumption, response time | Minimized energy consumption and response time | Minimize energy consumption, optimal offloading process |
| | GSP [118] | Joint optimization in mobile edge computing | Total energy consumed by devices and servers | Joint optimization, considering factors like offloading ratio, CPU speeds | Minimize total energy consumption |
| **Model-Free** | DDPG [119] | Collaborative MEC system with multi-users | Long-term energy consumption | Reduction in long-term energy consumption, optimize offloading, caching, resource allocation | Minimize long-term energy consumption, optimize resource allocation |
| | Meta Reinforcement | Adaptive task offloading strategy | Task processing delay, device energy consumption | Reduction in task processing delay, outperforms existing methods | Adapt to edge environment, reduce task processing delay |

| | | | | |
|---|---|---|---|---|
| | Learning [120] | | | |
| | TPDRTO [121] | Offloading computations considering task dependencies | Average energy consumption, time delay | Efficiently lowering the energy consumption and minimizing time delays for IoT devices | Optimize computation offloading, reduce energy consumption |
| | DQN [122] | Joint optimization in cloud-edge computation offloading | Average delay, average energy consumption, revenue | Comprehensive optimality on key indicators, outperforms baselines | Joint optimization of delay, energy consumption, and revenue |
| | Post-Decision State (PDS) Learning [123] | Offline value iteration and reinforcement learning | Long-term system cost | Improved edge computing performance, address energy harvesting challenges | Incorporate renewable energy, optimize offloading and autoscaling |
| **Hybrid** | Distributed Deep Learning [124] | Near-optimal computational offloading decisions | Overall energy consumption | Fast convergence, significant reduction in overall energy consumption | Find near-optimal offloading decisions, reduce overall energy consumption |
| | Deep Learning-based Offloading [125] | Optimal offloading scheme based on energy and performance | Energy consumption, performance constraints | Outperforms current approaches in meeting both energy and performance constraints | Compute optimal offloading scheme based on energy and performance |
| | Blockchain and DRL [126] | Energy-aware task scheduling and offloading | Consumable energy, QoS | 50% better energy efficiency, improved QoS | Enable energy-aware task scheduling, improve reliability |
| | DRL Algorithm [127] | Power minimization in cloud-edge-end collaboration | Power consumption | Superior power efficiency, quick convergence to a stable state | Minimize power consumption, optimize task offloading |

## 6. Conclusions

In this paper, we analyzed the architecture, variables, and decision-making algorithms involved in application offloading in IoT-based edge cloud environments focused on smart energy grid decentralization. Our study results show the urgent need to enhance the energy efficiency of cloud offloading and edge computing, especially concerning the specific problems of the smart grid and the transition towards renewable energy. The consideration of applications virtualization and microservices organization tailored to the IoT energy metering devices is a practical and forward-looking approach. The computing continuum organization from edge to fog and cloud can improve the service quality and significantly save bandwidth latency in the complex world of IoT-based energy management applications.

While current decentralized systems and offloading processes show potential, we also highlight their drawbacks, including complexity, high initial costs, and ongoing challenges with integration and security. The integration of the Smart grid architectures, known for their layered approach, with edge fog cloud computational resources organization presents promising solutions, however, their use requires careful consideration of smart grid functional, operational, and organizational requirements to ensure optimal usage.

There is a strong need for in-depth examination of the cloud- fog - edge architecture in the context of smart grid decentralization to maximize benefits and effectively address the challenges of renewable energy integration. As the amount of data generated by the smart grid IoT devices significantly grows, edge offloading and edge AI will be critical for enabling real-time response in emergency situations that may affect the grid resilience and at the same time, will help in addressing challenges related to limited network bandwidth and increased latency which affects decision making. In this context, we have analyzed and compared decision making algorithms based on metaheuristics and model-free optimization techniques like reinforcement learning and distributed deep learning for offloading.

These algorithms are fundamental for improving the overall performance of workload offloading in smart grid scenarios, requiring careful consideration of various decision-making criteria from energy and non-energy fields. In future work, researchers in computer science and the smart grid should focus on validating these algorithms through practical experiments in smart grid pilots, considering different deployment configurations, various computing resources, and distributed energy assets.

Table 5 presents strengths, weaknesses, opportunities, and threats (SWOT) for incorporating cloud offloading into smart grids. In our opinion, future research should focus on more exploration and innovation to tackle the weaknesses and threats, addressing relevant challenges of smart grid decentralization and IoT adoption, while considering emerging resources decentralization trends and AI advancement to continuously improve the decision-making strategies. Synergic efforts from energy, IoT, and AI domains are important for the smart grid to increase efficiency through integrating the decentralized renewable energy sources and creating sustainable and resilient future energy systems that meet the demands of customers delivering personalized and context-aware energy services.

**Table 5.** SWOT analysis of edge cloud offloading for smart grids.

| Strengths | Weaknesses | Opportunities | Threats |
| --- | --- | --- | --- |
| Offloading enhances overall efficiency by leveraging remote servers for computational tasks | Implementation of offloading architectures may face complexities, including integration issues with smart grid. | Technological trends in energy, IoT and AI create opportunities to improve offloading capabilities in smart grid | Security and privacy threats, requiring robust measures to prevent unauthorized access |
| Allows for real-time data processing, reducing latency and improving response time | Initial costs for deploying edge cloud offloading solutions in smart grid can be high | AI based development of optimization algorithms can improve edge cloud offloading strategies | Bandwidth and edge devices limitations may affect the effectiveness of offloading |
| Offloading architectures can improve grid resilience and energy security | Some energy applications requirements or constraints make then unsuitable for offloading | IoT adoption in smart grid offers opportunity to create new energy management applications. | Network stability can impact performance and reliability |
| Edge AI for optimal performance and decentralized energy services delivery | Distributing tasks across different smart grid layers affect coordination and edge-based orchestration | Cost-effective hardware solutions for edge and fog can increase the adoption of edge cloud offloading | Challenges in edge offloading deployments and data interoperability issues |
| Carbon saving to the integration of renewable on the far edge of the grid | Some computational architectures lack hierarchical distribution making challenging the smart grid integration | Smart grid architecture provides opportunity for customized solutions considering edge-fog-cloud distribution | Lack of validation in smart grid pilot to show the effectiveness of edge cloud offloading |

**Acknowledgments:** This work has been conducted within the HEDGE-IoT project grant number 101136216 funded by the European Commission as part of the Horizon Europe Framework Programme.

**Conflicts of Interest:** The authors declare no conflict of interest.